\documentclass[pdflatex,sn-mathphys-num]{sn-jnl}

\usepackage{graphicx}%
\usepackage{multirow}%
\usepackage{amsmath,amssymb,amsfonts}%
\usepackage{amsthm}%
\usepackage{mathrsfs}%
\usepackage[title]{appendix}%
\usepackage{xcolor}%
\usepackage{textcomp}%
\usepackage{manyfoot}%
\usepackage{booktabs}%
\usepackage{algorithm}%
\usepackage{algorithmicx}%
\usepackage{algpseudocode}%
\usepackage{listings}%
\usepackage{siunitx}
\usepackage{dcolumn}
\usepackage[normalem]{ulem}
 
\def\u#1{_{\rm #1}}
\newcommand{\enquote}[1]{``#1''}
\newcommand{\ket}[1]{| #1 \rangle}
\newcommand{\bra}[1]{\langle #1 |}
\newcommand{\ketbra}[2]{| #1 \rangle \langle #2 |}
\newcommand{\expect}[1]{\langle #1 \rangle}

\def\vac{\ket{{\rm vac}}}
\def\tr{{\rm Tr}}
\def\g2{g^{(2)}}

\DeclareRobustCommand{\erase}{\bgroup\markoverwith{\textcolour{red}{\rule[.5ex]{2pt}{0.4pt}}}\ULon}

\theoremstyle{thmstyleone}%
\theoremstyle{thmstyletwo}%

\theoremstyle{thmstylethree}%

\begin{document}

\title[Article Title]{Experimental entanglement swapping through single-photon $\chi^{(2)}$ nonlinearity}


\author*[1]{\fnm{Yoshiaki} \sur{Tsujimoto}}\email{tsujimoto@nict.go.jp}

\author[1]{\fnm{Kentaro} \sur{Wakui}}

\author[1]{\fnm{Tadashi} \sur{Kishimoto}}

\author[2]{\fnm{Shigehito} \sur{Miki}}

\author[2]{\fnm{Masahiro} \sur{Yabuno}}

\author[1,2]{\fnm{Hirotaka} \sur{Terai}}

\author[1]{\fnm{Mikio} \sur{Fujiwara}}

\author[1]{\fnm{Go} \sur{Kato}}

\affil[1]{\orgname{National Institute of Information and Communications Technology~(NICT)}, \city{Koganei}, \state{Tokyo} \postcode{184-8795}, \country{Japan}}

\affil[2]{\orgname{National Institute of Information and Communications Technology~(NICT)}, \city{Kobe}, \state{Hyogo} \postcode{651-2492}, \country{Japan}}


\abstract{
In photonic quantum information processing, quantum operations using nonlinear 
photon-photon interactions are vital for implementing two-qubit gates and enabling 
faithful entanglement swapping. 
However, due to the weak interaction between single photons, the all-photonic
realization of such quantum operations has remained out of reach so far.
Herein, we demonstrate an entanglement swapping using sum-frequency generation
between single photons in a $\chi^{(2)}$-nonlinear optical waveguide.
We show that a high signal-to-noise ratio~(SNR), stable sum-frequency-generation-based
entanglement heralder with an ultralow-dark-count superconducting single-photon detector 
can satisfy the unprecedented SNR requirement indispensable for the swapping protocol.
Furthermore, the system clock is enhanced by utilizing ultrafast telecom entangled 
photon-pair sources that operate in the GHz range.
Our results confirm a lower bound 0.770(76) for the swapped state’s fidelity, surpassing
the classical limit of 0.5 successfully. Our findings highlight the strong potential of
broadband all-single-photonic nonlinear interactions for further sophistication in
long-distance quantum communication and photonic quantum computation.
}

\maketitle

\section*{INTRODUCTION}\label{sec1}
Nonlinear interactions between independent single photons are key to the advancement of
photonic quantum information processing~\cite{OBrien2009-bd,10.1063/1.5115814}. 
In the past few decades, linear optical elements have been commonly adopted to configure such quantum information processing systems~\cite{Knill2001,PhysRevA.63.030301,PhysRevA.64.062311,PhysRevA.65.012314}. 
While these elements involve the use of general optical components, they require a large number of single-photon sources and detectors as the system sizes increase. 
The configuration of such systems may be dramatically simplified once the nonlinear photon-photon interaction is unlocked. For example, a complete Bell-state measurement~\cite{PhysRevLett.86.1370}, 
faithful entanglement swapping~\cite{PhysRevLett.106.120403} and even universal quantum computation~\cite{PhysRevLett.120.160502} can be conducted without
the need for complex ancillary systems by exploiting $\chi^{(2)}$ interaction between single photons. 

In the entanglement swapping protocol, as outlined in~\cite{PhysRevLett.106.120403}, a qualitative distinction arises between sum-frequency generation (SFG)-based Bell-state analyzers~(SFG-BSA) and those based on linear optics when entangled photon pairs are generated probabilistically. Specifically, while the quantum states heralded by linear-optical BSAs
become mixed states, the states heralded by SFG-BSAs remain close to maximally entangled states. These high-fidelity entangled states can be directly employed in various quantum protocols without the need for postselection.
In pursuit of this goal, several experimental attempts have been undertaken to observe SFG between single photons, including SFG between a heralded single photon and weak coherent light~\cite{Guerreiro2013}, as well as between independent heralded single photons~\cite{PhysRevLett.113.173601}. 
However, quantum operations have remained challenging mainly
due to the dark count noise of the single-photon detector comparable to the SFG
signal. Consequently, although there are several feasibility studies~\cite{PhysRevA.98.023836,PhysRevLett.123.250505}, no quantum operations using SFG between genuine single photons have been achieved thus far.

In this study, we overcome the signal-to-noise ratio~(SNR) problem and demonstrate an original entanglement swapping experiment based on SFG between two independent color-distinct single photons. 
We develop a high SNR and stable SFG-BSA unit, which performs the Bell-state measurement in the polarization degree of freedom with the help of $\chi^{(2)}$ nonlinearity. Our experimental
setup features a long periodically-poled lithium niobate waveguide~(PPLN/W)~\cite{Kishimoto:16} 
within a stable Sagnac interferometer for the SFG-BSA, and a ultralow-dark-count-rate superconducting nanowire single photon detector~(SNSPD) optimized for near-infrared detection~\cite{Miki:17}. 
Unlike systems that use Mach-Zehnder interferometers~(MZIs), an active stabilization 
is not required in our SFG-BSA unit, enduring weeks-order long-term data acquisition.
Moreover, we employ SPDC-based high-repetition-rate telecom entangled photon-pair sources~(EPSs)~\cite{Wakui:20} tunable in the GHz range to boost the generation rate of input entangled photon pairs. 

First, we performed a quantum-teleportation experiment to assess our SFG-BSA unit. We prepared
a weak coherent pulse at the single-photon level and an entangled photon pair as an input state. 
Then, we confirmed the SFG-BSA high-fidelity operation by teleporting the polarization state of
the weak coherent pulse using the detection signal of the SFG photon as a heralder.
Subsequently, we conducted the SFG-based entanglement swapping experiment using two entangled photon pairs and the SFG-BSA. To our knowledge, this study is the first to realize a quantum
operation using SFG between genuine single photons.

\section*{RESULTS}
\label{sec2}
\subsection*{The SFG-BSA configuration}
\label{subsec1}

\begin{figure}[t]
 \begin{center}
  \includegraphics[width=\columnwidth]{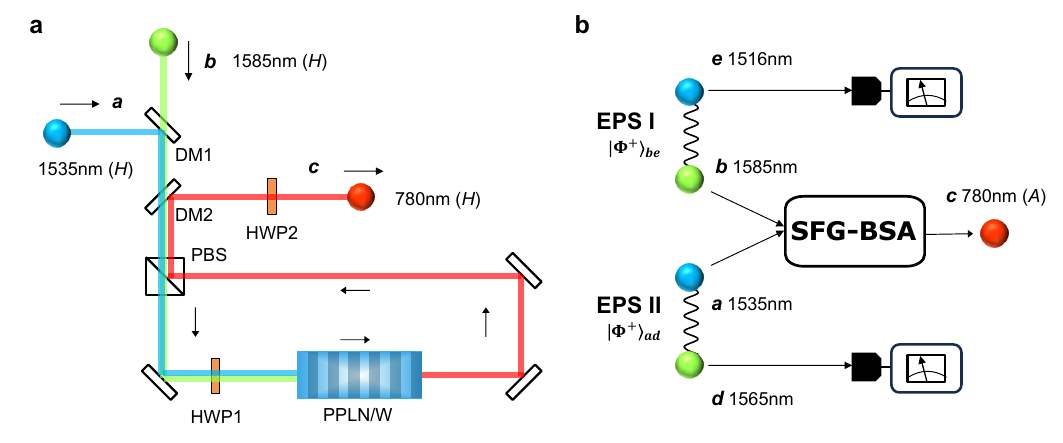}
  \caption{{\bf SFG-BSA and SFG-based entanglement swapping.} {\bf a,}~Schematic of the SFG-BSA, 
  showing the operation for two $H$-polarized input photons in mode $a$ and $b$, which are combined into 
  a single spatial mode by dichroic mirror 1~(DM1). They transmit through a polarization 
  beamsplitter~(PBS). Then, they flip to $V$ polarization by half waveplate 1~(HWP1). At the PPLN/W, 
  the $V$-polarized photons in mode $a$ and $b$ are converted to a $V$-polarized single photon in mode $c$ via 
  the SFG process. Finally, the SFG photon is extracted by DM2, and its polarization is flipped 
  back to $H$-polarization by HWP2. {\bf b,}~Schematic of the SFG-based entanglement swapping. 
  Detection of the $D$-/$A$-polarized SFG photon in mode $c$ at \SI{780}{nm} heralds a creation of 
  entanglement between the photons in mode $d$ and $e$.
  \label{fig1:QPC}} 
 \end{center}
\end{figure}

Fig.~\ref{fig1:QPC}{\bf a} shows a schematic of the SFG-BSA setup. A type-0 PPLN/W is incorporated
in a Sagnac interferometer to extract two of the four Bell states in polarization degree of freedom. It should be
noted that this degree of freedom was preferred over the time degree of freedom, as the latter incurs photon losses by configuring the
measurement system with a passive Franson interferometer~\cite{PhysRevLett.82.2594}. 
In contrast to MZI-based configurations~\cite{Hamel2014}, our SFG-BSA does not require phase stabilization between horizontally~($H$-) and vertically~($V$-) polarized photons. 
Thus, it can be viewed as a time-reversal of the polarization-entangled photon-pair generation using a Sagnac interferometer with a PPLN/W~\cite{Wakui:20,Yamazaki2022,Tsujimoto2018}.  
In this configuration (Fig.~\ref{fig1:QPC}{\bf a}), no SFG photon is produced if the two input photons 
are $H$- and $V$-polarized, respectively. However, if both input 
photons are $V$-($H$-)polarized, they propagate clockwise~(counterclockwise) through the interferometer 
to produce $V$-($H$-)polarized SFG photons, respectively. 

This can be further discussed through the interaction Hamiltonian: 

\begin{equation}
\hat{H}=i\hbar\chi(\hat{a}_H\hat{b}_H\hat{c}^\dagger_H+\hat{a}_V\hat{b}_V\hat{c}^\dagger_V)+\mathrm{H.c.}, 
\label{eq1:SFG}
\end{equation}
where $\hat{a}_k$ and $\hat{b}_k$ are annihilation operators of
single photons with polarization $k\in\{H, V\}$ in the input mode $a$ and $b$, respectively,  
$\hat{c}_k$ is an annihilation operator of a $k$-polarized single photon in the output mode $c$,
H.c. is the Hermitian conjugate and $\hbar$ is the Dirac's constant. 
Here, $\chi\in\mathbb{R}$ is a coupling constant that includes the nonlinear susceptibility of PPLN/W, and it 
is assumed to be independent of the polarization of the input photons. 
Additionally, we assume that the sum of the photon numbers in the input mode $a$ and $b$ does not exceed two,
and we consider only the lowest order of $\chi$. 
These assumptions are made to describe the ideal operation of the SFG-BSA. 
The more realistic situations including polarization dependency of $\chi$ and multiphoton inputs are considered in Theoretical analysis in Methods section. 
In the case where one photon is generated in mode $c$ as a success event, the corresponding Kraus operator is given by

\begin{equation}
\hat{K}:=\sqrt{\eta\u{SFG}}(\ket{H}_c\bra{HH}_{ab}+\ket{V}_c\bra{VV}_{ab}),
\label{eq2:QPC}
\end{equation}
where $\eta\u{SFG}:=(\chi \tau)^2$ and
$\ket{k}_a:=\hat{a}^\dagger_{k}\vac$ for $k\in\{H,V\}$ and so on. 
Here, $\tau$ is the travel time of the input photons through the nonlinear medium and
$\vac$ is a vacuum state. 
By detecting a diagonally~($D$-) or anti-diagonally~($A$-) polarized SFG photon in mode $c$ 
(defined by  $\ket{D}_c:=(\ket{H}_c+\ket{V}_c)/\sqrt{2}$ and $\ket{A}_c:=(\ket{H}_c-\ket{V}_c)/\sqrt{2}$), 
the projection on a Bell state
$\ket{\Phi^{+}}_{ab}:=\frac{1}{\sqrt{2}}(\ket{HH}_{ab}+\ket{VV}_{ab})$ 
or $\ket{\Phi^{-}}_{ab}:=\frac{1}{\sqrt{2}}(\ket{HH}_{ab}-\ket{VV}_{ab})$ 
will be respectively performed, which corresponds to a quantum parity check. 

Compared to a linear-optical BSA~\cite{PhysRevLett.80.3891,PhysRevA.63.030301,PhysRevA.64.062311, Halder2007,Jin2015, Tsujimoto2018}, 
the SFG-BSA offers the following unique benefits. First, it aids faithful entanglement swapping based on probabilistic photon pair sources such as SPDC~\cite{PhysRevLett.106.120403}, in which
detection of a SFG photon deterministically heralds the creation of an entangled photon pair.
This makes it well suited for a long-distance loophole-free Bell test and device-independent quantum key distribution~\cite{PhysRevLett.98.230501}. 
Interestingly, this principle allows tolerance against optical loss for entanglement swapping~(details are given in the next section). 
Second, by adding another SFG-BSA for odd-parity inputs, it enables a complete Bell-state measurement that distinguishes all four Bell states without
employing ancillary systems~\cite{PhysRevLett.86.1370}. 
Finally, it allows entangling operations between different-color photons, which is useful for building multi-user entanglement networks~\cite{PhysRevLett.123.250505}. 

Fig.~\ref{fig1:QPC}{\bf b} describes the entanglement swapping experiment based on the SFG-BSA. 
EPS~I and EPS~II generate two maximally entangled photon pairs
$\ket{\Phi^+}_{ad}\otimes\ket{\Phi^+}_{be}$ and the SFG-BSA is performed on the photons in mode $a$ and $b$, yielding $_c\bra{D(A)}\hat{K}\ket{\Phi^+}_{ad}\otimes\ket{\Phi^+}_{be}\propto\ket{\Phi^{+(-)}}_{de}$, 
respectively. The overall success probability is
$\sum_{l\in \{D,A\}}|_c\bra{l}\hat{K}\ket{\Phi^+}_{ad}\otimes\ket{\Phi^+}_{be}|^2=\eta\u{SFG}/2$. 
The experimental success probability is $\eta\u{SFG}/4$ considering that only one of the
$D$- or $A$-polarized portions is measured. 

\subsection*{Robustness of SFG-BSA against optical losses}
\label{subsec2}

To see the loss tolerance of the SFG-based entanglement swapping with probabilistic sources, we consider error events of the lowest order, where two photon pairs are generated by EPS~I and one photon pair is generated by EPS~II. In the presence of high channel loss, the dominant error event is the loss of a single photon among the three photons transmitted through the channel. This type of event can be categorized into two distinct types. First, one photon from EPS~I is lost, and one photon from each of EPS~I and EPS~II successfully arrives at the BSA, which results in a coincidence detection. Second, one photon from EPS~II is lost, and a coincidence detection is induced by two photons from EPS~I. Although the first error event commonly occurs in the linear-optical BSA and SFG-BSA, the second error event can be excluded using the SFG-BSA. The reason is that the SFG occurs only when at least one photon arrives from each EPS, which results in an increase in the robustness of the SFG-BSA against optical loss. Let $\gamma^2$ be the 
respective photon pair generation probability of EPS~I and EPS~II, and let $t$ be the transmittance
of each channel. Then, the probabilities of the first and second error events
are given by $2\gamma^6t^2(1-t)$ and $\gamma^6t^2(1-t)$, respectively. 
Thus, 1/3 of the error events are excluded using SFG-BSA. More detailed analysis is given in Supplementary Note 1 and 2.

\subsection*{SFG-BSA unit}
\label{subsubsec1}

The SFG-BSA unit consists of an in-house type-0 MgO-doped PPLN ridge waveguide~(PPLN/W3) with 
free space coupling (length: \SI{6.3}{cm}). 
Details of device fabrication are given in Supplementary Note 5. 
To estimate the single-photon SFG efficiency, we used
coherent light pulses centered at 1535 and \SI{1585}{nm}. These pulses were generated according
to the difference-frequency generation~(DFG) at EPS~I and II in Fig.~\ref{fig2:Experiment}
with the pulsed pump at \SI{775}{nm} and the two additional continuous wave lasers at 1565 and \SI{1516}{nm}, respectively. Their coupling efficiencies to the PPLN/W3 
were measured to be 0.77 and 0.89, respectively, indicating a large mode matching between the
input-photon and waveguide modes. 
The estimated spatial profiles of photons in mode $a$, $b$ and $c$ are shown in Supplementary Note 6.
By measuring the average power of the DFG pulses and the count
rate of the SFG photons, the internal SFG conversion efficiency for single-photon inputs can be
estimated as 
\begin{equation}
\eta\u{SFG}=\frac{C\u{SFG}}{\eta_t\eta_df}\times \frac{hcf}{P_a\lambda_a}\times \frac{hcf}{P_b\lambda_b},
\label{eq3:SFGeff}
\end{equation}
where $c$ is the speed of light, $h$ is the Planck’s constant, $\eta_t$ and $\eta_d$ are the
transmittance of the system and the quantum efficiency of the SNSPD used to detect the SFG photons, respectively. 
In addition, $P_{a(b)}$ and $\lambda_{a(b)}$ are the average power per mode and wavelength of the DFG pulses
in mode $a$ and $b$, respectively. $C\u{SFG}$ is the count rate of the SFG photons, and $f$=\SI{1.0}{GHz} 
is the repetition rate of the DFG pulses. Using the experimental parameters summarized in
Table~\ref{table1:SFGeff}, we obtained $\eta^H\u{SFG}=2.31\times10^{-8}$ and 
$\eta^V\u{SFG}=2.35\times10^{-8}$ for $H$- and $V$-polarized inputs, respectively.

\subsection*{SFG-based quantum teleportation}
\label{subsubsec2}

To assess our SFG-BSA unit, we performed a quantum teleportation experiment. The input light to be teleported was
generated by performing DFG at EPS~II. We prepared $H$-, $A$-, and right circularly~($R$-) polarized weak coherent
light as input states, and we set the average photon number at 0.95 per pulse. Then, we reconstructed the density
matrices of the input states by conducting quantum state tomography~\cite{PhysRevA.64.052312}. 
Fig.~\ref{fig3:Teleportation}{\bf a} shows an example of the density matrix for the $A$-polarized 
state~(the density matrices of the $H$- and $R$-polarized input states are shown in Supplementary Note 3).
The fidelities of the $H$-, $A$- and $R$-polarized states are
$\bra{H}\hat{\rho}_b\ket{H}=0.97050(9)$, $\bra{A}\hat{\rho}_b\ket{A}=0.99735(3)$ and $\bra{R}\hat{\rho}_b\ket{R}=0.97500(8)$, 
respectively, and here $\ket{R}:=(\ket{H}+i\ket{V})/\sqrt{2}$. The fidelity errors correspond to the standard
deviations with the assumption of Poisson statistics for the photon counts. In addition, we prepared an entangled
photon pair using EPS~I, whose density matrix is shown in Fig.~\ref{fig3:Teleportation}{\bf b}. 
The fidelity to the ideal state $\ket{\Phi^+}$ is $\bra{\Phi^+}\hat{\rho}_{ad}\ket{\Phi^+}=0.9163(4)$. 
We input the weak coherent light in mode $b$ and the photon in mode $a$ into the SFG-BSA. Under the condition
that a $D$-polarized SFG photon was detected by D1, we performed the quantum state tomography on the photon in mode $d$~(see Fig.~\ref{fig2:Experiment}). 
Fig.~\ref{fig3:Teleportation}{\bf c} and \ref{fig3:Teleportation}{\bf d} show the density matrix of the teleported
state and the raw detection counts for the $A$-polarized input, respectively. The fidelity is 0.890(30), and the 
measurement time is \SI{13}{h} for each basis state. For $H$- and $R$-polarized inputs, the fidelities are 0.893(28) 
and 0.840(39), respectively. These values significantly exceed the classical limit of 2/3~\cite{PhysRevLett.74.1259}, 
indicating the high fidelity of the SFG-BSA for inputs at the single photon level. It is worth noting that the results
in Ref.~\cite{PhysRevLett.86.1370} were reported as quantum teleportation using SFG with $10^{10}$ input photons on average. 
Nevertheless, these results can be rather understood as a quantum frequency conversion~\cite{Kumar:90, Tanzilli2005, Ikuta2011} 
in the weak pump regime, in contrast to the present quantum-teleportation experiment with single-photon-level inputs. 
This is because in the situation of Ref.~\cite{PhysRevLett.86.1370}, the input light is regarded as the pump light for the quantum frequency conversion. Hence, the quantum-teleportation-like behavior is observed under the condition that the conversion efficiency
of the quantum frequency conversion can be approximately regarded as proportional to the average intensity of the input light. In the strong-pump regime, where the above approximation does not hold, the polarization state is no longer transferred.
A detailed discussion is given in Supplementary Note 4.

\subsection*{SFG-based entanglement swapping}
\label{subsubsec3}

Fig.~\ref{fig2:Experiment} depicts the experimental setup for
entanglement swapping. We prepared two initial entangled photon pairs
$\hat{\rho}_{ad}$ and $\hat{\rho}_{be}$ as input states. Their
density matrices are shown in Figs.~\ref{fig4:Swapping}{\bf a} 
and \ref{fig4:Swapping}{\bf b}, respectively. The fidelities are 
$\bra{\Phi^+}\hat{\rho}_{ad}\ket{\Phi^+}=0.9079(4)$ and 
$\bra{\Phi^+}\hat{\rho}_{be}\ket{\Phi^+}=0.8775(6)$. The photons in mode
$a$ and $b$ were combined by a volume holographic grating and
fed to the SFG-BSA. When an $A$-polarized SFG photon at \SI{780}{nm} 
is detected by the SNSPD~(D1), the both output ports of the fiber-based 
polarizing beamsplitter were simultaneously measured using SNSPDs D2, D3, D4, and D5. The low-noise detection of the SFG photon in mode $c$ was enabled by D1, which possesses a high quantum efficiency~($\eta_d$=\SI{85}{\%}) and an ultralow dark count
rate~($R_d=$\SI{0.15}{Hz}). The typical quantum efficiency of D2, D3, 
D4, and D5 was \SI{75}{\%}. Fig.~\ref{fig4:Swapping}{\bf c} shows an
example of the two-fold coincidence between D1 and D2. We observe a
signal peak corresponding to the coincidence detection between photons in mode
$c$ and $d$. Background counts are primarily caused by the coincidence
between the dark count at D1 and the detection of the photon in mode $d$ at D2. 
We used a coincidence window of $\tau\u{w}=$\SI{448}{ps}, corresponding
to seven bins around the signal peak, shown by the red bars in
Fig.~\ref{fig4:Swapping}{\bf c}. The three-fold coincidence histogram
among D1, D2, and D4 with bin widths $\tau\u{w}$ is shown in
Fig.~\ref{fig4:Swapping}{\bf d}. We clearly identify a signal peak
corresponding to $P_{VV}:=\bra{VV}\hat{\rho}_{de}\ket{VV}$, with a
measured SNR equal to $5.57\pm$1.48. We performed Z-basis measurements
that distinguish the $H/V$ polarization and X-basis measurements that
distinguish the $D/A$ polarization on the photons in mode $d$ and $e$. 
Each measurement~(30-min duration) was repeated 452 times, resulting in
an overall measurement time of \SI{226}{h} for each basis state. 
The coincidence counts of the swapped state for eight basis states are
shown in Figs.~\ref{fig4:Swapping}{\bf e} and \ref{fig4:Swapping}{\bf f}. 
The polarization correlation corresponding to the target swapped state 
$\ket{\Phi^-}_{de}:=(\ket{HH}_{de}-\ket{VV}_{de})/\sqrt{2}=(\ket{DA}_{de}+\ket{AD}_{de})/\sqrt{2}$ 
is clearly identified. We evaluated the quality of the swapped state
by the visibilities for Z-basis~($V\u{Z}:=\expect{\hat{Z}_d\hat{Z}_e}$) 
and X-basis~($V\u{X}:=-\expect{\hat{X}_d\hat{X}_e}$), where $\hat{Z}:=\ketbra{H}{H}-\ketbra{V}{V}$ 
and $\hat{X}:=\ketbra{H}{V}+\ketbra{V}{H}$ are Pauli operators. We observed high visibilities 
of $V\u{Z}=0.833(92)$ and $V\u{X}=0.706(121)$. Although these
parameters are not sufficient to reconstruct the complete density matrix, 
they help to estimate the lower bound of the fidelity
$F \geq F\u{low} = (V\u{Z}+V\u{X})/2$~\cite{PhysRevA.65.042314,PhysRevLett.106.110503}. 
The experimental value of $F\u{low}$ was found to be equal to 0.770(76), 
which confirms that the swapped state is strongly entangled.

\section*{DISCUSSION}
\label{sec3}

We compare the performance of the presented approach based on the SFG-BSA with existing linear optical approaches. As outlined in the theoretical proposal~\cite{PhysRevLett.106.120403}, a key application of the SFG-BSA is the heralded generation of entangled photon pairs using probabilistic photon sources, a method wherein the preparation of an entangled photon pair is conditioned on the detection of auxiliary photons. Specifically, upon the detection of a heralding signal, the entangled photon pair is deterministically prepared and can subsequently be utilized in various quantum protocols.
The linear optics-based approach~\cite{PhysRevA.67.030101} involves the preparation of three pairs of twin photons~(six photons in total), utilizing four of these photons to herald the generation of an entangled photon pair. As discussed in previous studies~\cite{PhysRevA.67.030101,PhysRevLett.106.120403}, there exists a trade-off between the success probability and the fidelity of the heralded state. In this context, we compute the maximum success probability that ensures a fidelity greater than 0.9 using the equations provided in~\cite{PhysRevLett.106.120403}. Assuming a photon detection efficiency of 0.7, the success probability for the linear optical implementation is calculated to be $1.5\times 10^{-11}$. In comparison, using SFG-BSA, our theoretical model yields a success probability of $1.2\times10^{-11}$ with an SFG efficiency of $2.3\times10^{-8}$ and a photon detection efficiency of 0.7, which is nearly identical to the linear optical approach. Here, we assume that a complete Bell-state measurement is performed at the SFG-BSA. Furthermore, in regions with lower detection efficiency, the SFG-BSA becomes more efficient. For instance, at a photon detection efficiency of 0.5, the SFG-BSA exhibits a performance 16 times more efficient than the linear optics-based approach.

Another important application of the SFG-BSA is its use in the heralding scheme for a loophole-free Bell test with probabilistic photon sources. Upon detection of the heralding signal, the detection loophole induced by transmission losses can be effectively closed. In this context, linear optical BSA has been identified as one of the plausible schemes for achieving such a heralded Bell test~\cite{PhysRevA.84.010304, Tsujimoto_2020}. However, a qualitative difference exists between the quantum states heralded by linear optical BSA and those heralded by SFG-BSA. In a Bell test experiment, the correlation of the measurement outcomes is typically characterized by the parameter $S$, with the Bell-CHSH inequality stipulating that $|S|<2$ within the framework of local realism. According to~\cite{Tsujimoto_2020}, using linear optical BSA as a heralding scheme, the maximum achievable value of $S$ was shown to be 2.34. Furthermore, it was shown that a detection efficiency of at least 0.911 is required for both Alice's and Bob's detectors to observe a violation of the Bell-CHSH inequality. In contrast, our simulation results indicate that, when employing the SFG-BSA approach, the achievable value of $S$ approaches the Tsirelson bound of $2\sqrt{2}$~\cite{Cirelson1980-ly}. Additionally, we find that a detection efficiency of 0.68 is sufficient to observe the Bell-CHSH inequality violation, significantly reducing the stringent requirements for Alice and Bob's detection systems compared to the linear optical BSA approach. These results can be attributed to the fact that the fidelity of the quantum state heralded by the SFG-based BSA can approach unity, whereas that of the state heralded by the linear optical BSA is fundamentally limited to a maximum of 1/2 in the absence of postselection~\cite{PhysRevLett.106.120403}.

Finally, we examine the extent to which the efficiency of SFG must be improved for future applications. The SNR at the SFG photon detection achieved by our current system is insufficient to herald an entangled state capable of violating the Bell-CHSH inequality. However, theoretical simulations using experimental parameters from Tables~\ref{table1:SFGeff} and \ref{table2:Pnum} indicate that tripling the SFG efficiency would result in an adequate SNR to
herald the quantum state that exhibits a violation of the Bell-CHSH inequality. For the application to DIQKD, a further enhancement of the SFG efficiency by a factor of 50 would be required. This level of improvement is considered possible, given recent advancements in nonlinear resonators, which have demonstrated increases in efficiency by two to three orders of magnitude compared to PPLN/W~\cite{Zhu:21,Akin2024-cg}. (For further details on these simulations, see Supplementary Note 7.)

In conclusion, we have implemented a BSA using SFG between single photons and demonstrated the quantum teleportation and entanglement swapping. 
We have confirmed the high fidelity of the teleported/swapped states, which is an 
important milestone toward quantum information processing using single-photon $\chi^{(2)}$ nonlinearity. 
Toward practical applications, SFG efficiency improvements of several orders of magnitude are needed. Nevertheless, our simulation shows that even a several-fold improvement in the SFG efficiency is useful for the loophole-free Bell test. Recently, there has been notable progress in the research on the nonlinear devices. 
Combined with these technologies, our method provides a new avenue toward all-photonic quantum information processing using the single-photon $\chi^{(2)}$ nonlinearity.

\section*{Note added} During the preparation of the manuscript presented here, we learned of an experiment demonstrating quantum teleportation using single-photon SFG by Akin {\it et al.}~\cite{PhysRevLett.134.160802}.

\section*{METHODS}

\subsubsection*{Entangled photon pair sources}
We employed an electro-optic frequency comb generator to emit a
fundamental pulse to pump SPDC and realized a high-speed generation of
entangled photon pairs. The setups of EPS~I and EPS~II are shown on the
left side of Fig.~\ref{fig2:Experiment}. 
We generated a frequency comb centered at \SI{1550}{nm} with a 10-GHz 
repetition rate by electro-optic phase modulation on the narrow-band continuous wave laser
light~\cite{Sakamoto:07,Wakui:20,Jin2014,Tsujimoto:21}. We reduced it to 1/10 through
optical gating~\cite{Wakui2024}. We selected a 1.0-GHz repetition rate
such that the SNSPDs can resolve each pulse without saturating. Pump pulses at \SI{775}{nm} for SPDC were generated by SHG using a 1.0-cm-long PPLN/W. 
The SHG spectral bandwidth was measured to be 34-GHz full-width at half maximum~(FWHM) and the power was stabilized during the experiment. 
Each EPS consists of a 3.4-cm-long PPLN/W with a Sagnac
configuration~\cite{Wakui:20,Tsujimoto2018}. The pump power coupled to the PPLN/W 
was approximately \SI{1.0}{mW} for both the clockwise and counterclockwise
directions. We used volume holographic gratings with a 55-GHz FWHM
(i.e., in accordance with the phase-matching bandwidth of the SFG-BSA unit) 
to narrow the bandwidths of the signal and idler photons. 
The spectral distributions of the photons in mode $a$, $b$ and $c$ are shown in Supplementary Note 6.
Table~\ref{table2:Pnum} 
summarizes the average photon numbers per mode~($\mu$), Klyshko 
efficiencies~\cite{Klyshko_1980}~($\eta$), and the transmittance~($t$) in the optical
circuit before the SFG-BSA including the coupling efficiencies to the SFG-BSA. 
We employed a flip mirror~(not shown) just before the SFG-BSA to evaluate
the input quantum states from EPS~I and II. The detection rate of the entangled
photon pairs for each EPS was measured to be \SI{1.1}{MHz}.

\subsection*{Theoretical SFG efficiency}
The PPLN/W3 used for SFG possesses a crystal length of $L$=\SI{6.3}{cm} and 
a core diameter of 7.2\,$\mu$m$\times$8.0\,$\mu$m~\cite{Kishimoto:16}.
The normalized SHG efficiency was measured to be $\eta\u{SHG}=28\,\%\cdot \mathrm{W^{-1}}\cdot\mathrm{cm^{-2}}$ 
using a narrow-band continuous-wave laser light. The theoretical SFG conversion efficiency for single-photon 
inputs can be estimated using the following equation~\cite{PhysRevLett.106.120403,Guerreiro2013}:

\begin{equation}
\eta^{\mathrm{th}}\u{SFG}=\frac{\eta\u{SHG}}{2}\times\frac{hc}{\lambda}\times\frac{\Delta\hat{\nu} L}{\rm{tbp}}, 
\end{equation}
where $\Delta\hat{\nu}$ is the spectral acceptance of the crystal and $\rm{tbp}$ 
is the time-bandwidth product of the SFG photon. By substituting the 
experimental values of $\lambda=$\SI{1560}{nm}, $\Delta\hat{\nu}=2.48\times10^2$\,GHz$\cdot$cm 
and $\rm{tbp}=0.67$, we obtain $\eta^{\mathrm{th}}\u{SFG}=4.16\times10^{-8}$. 
The SFG efficiency for SPDC photons is estimated considering the overlap
integral among the spectral distributions of the input photons $\phi_{a(b)}(\lambda_{a(b)})$, 
and the normalized phase-matching function of PPLN/W3 $\eta(\lambda_a,\lambda_b)$ as~\cite{Guerreiro2013} 

\begin{equation}
\eta^{\mathrm{eff}}\u{SFG}=\eta^{\mathrm{th}}\u{SFG}\int\int \phi_a(\lambda_a)\phi_b(\lambda_b)\eta(\lambda_a,\lambda_b)d\lambda_ad\lambda_b.
\end{equation}
From the measured results, when $\phi_a(\lambda_a)$, $\phi_b(\lambda_b)$, $\eta(\lambda_a,\lambda_b)$ 
are calculated as Gaussians with FWHM of 0.31, 0.33, and \SI{0.080}{nm}, respectively, we obtain
$\eta^{\mathrm{eff}}\u{SFG}=2.42\times10^{-8}$, which is close to the experimental values of
$\eta^{H}\u{SFG}=2.31\times10^{-8}$ and $\eta^V\u{SFG}=2.35\times10^{-8}$.

\subsection*{Widths of coincidence windows}
The width of the coincidence window is determined according to the
timing jitters of the SNSPDs. The SFG photon in mode $c$ at \SI{780}{nm} is detected by D1, whose timing jitter is \SI{190}{ps}. The telecom photons are detected by D2, D3, D4 and D5, whose timing jitters are 48, 98, 64, and \SI{162}{ps}, respectively~(Fig.~\ref{fig2:Experiment}). The largest timing jitter among the three-fold coincidences is given according to the convolution of the timing jitters of D1, D3, D5 and the coherence time of SPDC photons of \SI{14}{ps} as \SI{269}{ps}. We used the coincidence window with $\tau\u{w}$=\SI{448}{ps}, which covers \SI{96}{\%} of the signal events.

\subsubsection*{Theoretical analysis of visibilities}
We discuss the validity of the experimentally obtained visibilities, 
using a realistic model and independently measured experimental parameters. 
We consider the dark counts in D1 and the multiple photons generated 
in EPS~I and EPS~II using the following approach. First, we consider the
events where a maximum of three photon pairs is produced between EPS~I 
and EPS~II as an input. Then, we perform the SFG operation after the photons
in mode $a$ and $b$ experienced system losses. We assume that the coupling constant 
$\chi$ in Eq.~(\ref{eq1:SFG}) is dependent on the inputs’ polarization, 
which is defined as $\chi_{H(V)}$ for $H$-($V$-) polarized input, respectively. 
In this model, a maximum of one SFG photon is produced, which is detected by
a threshold detector D1. Conditioned by the detection signal from D1, polarization
correlation measurements were performed using D2, D3, D4 and D5. The three-fold coincidence probability is given by $P^\mathrm{SFG}_{ij}(\theta_1,\theta_2)$ 
with $i,j\in\{H,V\}$. Here, Z-basis and X-basis are expressed by $(\theta_1,\theta_2)=(0,0)$ 
and $(\pi/4,\pi/4)$, respectively. For example, $P^\mathrm{SFG}_{HV}(\pi/4,\pi/4)$ 
corresponds to the coincidence probability where $D$- and $A$-polarized photons are
detected by D3 and D4 according to the heralding signal from D1, respectively. 
We also consider the event where the heralding signal from D1 is caused by the dark counts. Because the dark count probability at D1 is given by $R_d\times\tau_W$, 
the coincidence detection probability between the photons in mode $d$ and $e$ heralded by
the dark count at D1 is given by 
$R_d\tau_W(P^\mathrm{Acd}_{ij}(\theta_1,\theta_2)-P^\mathrm{SFG}_{ij}(\theta_1,\theta_2))$, 
where $P^\mathrm{Acd}_{ij}(\theta_1,\theta_2)$ is the accidental coincidence detection
probability between the photons in mode $d$ and $e$. The visibilities of the swapped state are calculated using the above coincidence probabilities as 
\begin{equation}
V\u{Z}^\mathrm{th}=\frac{P_{HH}(0,0)+P_{VV}(0,0)-P_{HV}(0,0)-P_{VH}(0,0)}{P_{HH}(0,0)+P_{VV}(0,0)+P_{HV}(0,0)+P_{VH}(0,0)}
\label{eq9:VZ}
\end{equation}
and
\begin{equation}
V\u{X}^\mathrm{th}=\frac{P_{HV}(\frac{\pi}{4},\frac{\pi}{4})+P_{VH}(\frac{\pi}{4},\frac{\pi}{4})-P_{HH}(\frac{\pi}{4},\frac{\pi}{4})-P_{VV}(\frac{\pi}{4},\frac{\pi}{4})}{P_{HV}(\frac{\pi}{4},\frac{\pi}{4})+P_{VH}(\frac{\pi}{4},\frac{\pi}{4})+P_{HH}(\frac{\pi}{4},\frac{\pi}{4})+P_{VV}(\frac{\pi}{4},\frac{\pi}{4})}, 
\label{eq10:VX}
\end{equation}
 where $P_{ij}(\theta_1,\theta_2):=P^\mathrm{SFG}_{ij}(\theta_1,\theta_2)(1-R_d\tau_W)+R_d\tau_WP^\mathrm{Acd}_{ij}(\theta_1,\theta_2)$.
 The detailed calculation and the theoretical model are presented
in Supplementary Note 1.

By substituting $\eta^H\u{SFG}=2.31\times10^{-8}$, $\eta^V\u{SFG}=2.35\times10^{-8}$, 
and the parameters in Tables~\ref{table1:SFGeff} and \ref{table2:Pnum}, 
the theoretically calculated visibility values are
$V^\mathrm{th}\u{Z}=0.78$ and $V^\mathrm{th}\u{X}=0.76$, respectively, 
which exhibit a good agreement with the experimentally observed visibilities
$V\u{Z}=0.833(92)$ and $V\u{X}=0.706(121)$. 

\subsection*{DATA AVAILABILITY}
The authors declare that the data supporting the findings of this study are available within the paper, its supplementary information files, and
Figshare~[https://doi.org/10.6084/m9.figshare.29919704].

\section*{ACKNOWLEDGEMENTS}
Y. T. thanks Rikizo Ikuta and Toshiki Kobayashi for helpful discussions. 
This work was supported by Japan Society for the Promotion of Science (JP18K13487, JP20K14393, JP22K03490) and R\&D of ICT Priority Technology Project (JPMI00316). 


\section*{AUTHOR CONTRIBUTIONS}
Y.T. conceived the idea, conducted the experiment with the assistance of K.W.,
and performed the simulation with the assistance of G.K.. 
K.W., G.K., and M.F. provide valuable advice and discussions on
theory and experiment. 
T.K. fabricated PPLN/W3 used for the SFG-BSA unit. S.M., M.Y., and H.T. 
developed the SNSPD system. 
Y.T. wrote the manuscript with the help of all the other authors.

\section*{COMPETING INTERESTS}
The authors declare no competing interests.

\begin{table}[h]
\caption{{\bf Parameters of the SFG-BSA unit}}\label{table1:SFGeff}%
\begin{tabular}{@{}llllll@{}}
\toprule
& $C\u{SFG}$ & $\eta_t$ &  $\eta_d$ & $P_a$ & $P_b$\\
\midrule
 $H$ & \SI{2.54}{MHz} & 0.43 & 0.85 & \SI{80}{nW} & \SI{61}{nW} \\
 $V$ & \SI{1.94}{MHz} & 0.40 & 0.85 & \SI{70}{nW} & \SI{56}{nW}  \\
\botrule
\end{tabular}
\end{table}

\begin{table}[h]
\caption{{\bf Parameters of ESP~I and EPS~II.}}\label{table2:Pnum}%
\begin{tabular}{@{}llllllll@{}}
\toprule
 & \multicolumn{3}{c}{EPS I} & & \multicolumn{3}{c}{EPS II} \\
\midrule
      & $\mu_{1}$ & $\eta_{1}$ & $t_{1}$& & $\mu_{2}$ & $\eta_{2}$ & $t_{2}$ \\
  $H$  & 0.060 & 0.097 & 0.44 & & 0.080 & 0.070 & 0.56 \\
  $V$  & 0.050 & 0.11 & 0.48 & & 0.061 & 0.10 & 0.57 \\
\botrule
\end{tabular}
\end{table}

\begin{figure}[t]
 \begin{center}
  \includegraphics[width=\columnwidth]{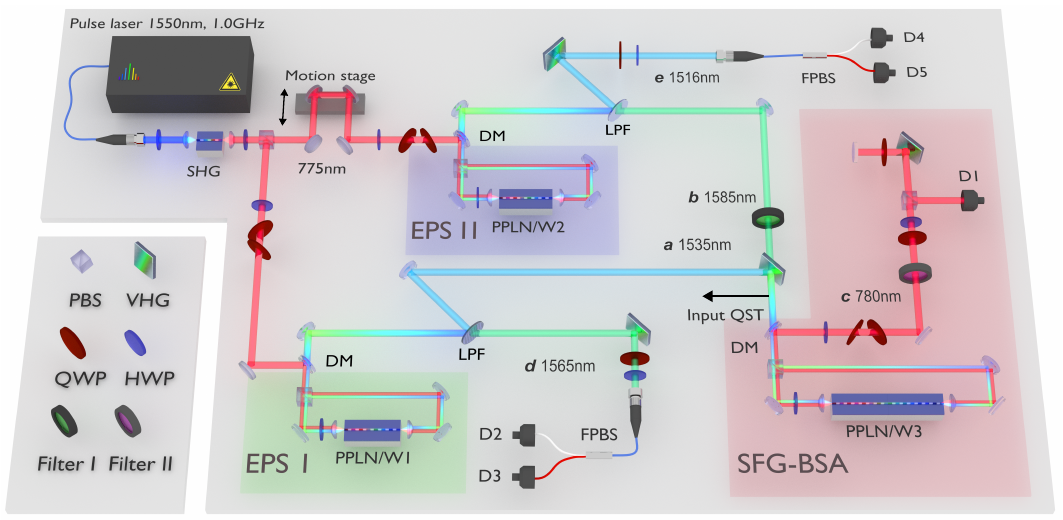}
  \caption{{\bf Experimental setup for the SFG-based entanglement swapping.} 
  Pump pulses centered at \SI{775}{nm} with a 1.0-GHz repetition rate is prepared by 
  second harmonic generation~(SHG) of 
  the electro-optic comb centered at \SI{1550}{nm} and used to pump EPS~I and II. 
  Each EPS consists of a PPLN/W in a Sagnac interferometer with a polarizing beamsplitter~(PBS). 
  The signal and idler photons at telecom wavelengths 
  are divided into different spatial modes according to low pass filters~(LPFs). 
  The photon in mode $a$ at \SI{1535}{nm} and the photon in mode $b$ at \SI{1585}{nm} are narrowed by a volume 
  holographic grating~(VHG) and band pass filter (Filter~I), respectively, and they are 
  fed into the SFG-BSA unit. A flip mirror (not shown) just before the SFG-BSA is used 
  to perform the quantum state tomography~(QST) on the input quantum states. The output SFG 
  photon in mode $c$ at \SI{780}{nm} is extracted by a dichroic mirror~(DM) and passes through the band pass filter~(Filter~II) and is 
  diffracted twice by a VHG. The polarization of the SFG photon is projected on the $A$-polarization 
  by means of a quarter waveplate~(QWP), half waveplate~(HWP), and PBS. 
  The photon-detection signal from the SNSPD~(D1) is used as the start signal of a time-to-digital 
  converter~(not shown). The photons in mode $d$ and $e$ are diffracted by VHGs. The polarization correlation 
  of the swapped state $\hat{\rho}_{de}$ is evaluated by QWPs, HWPs, and fiber-based PBSs~(FPBSs) 
  followed by SNSPDs (D2-D5).
  \label{fig2:Experiment}} 
 \end{center}
\end{figure}

\begin{figure}[t]
 \begin{center}
  \includegraphics[width=\columnwidth]{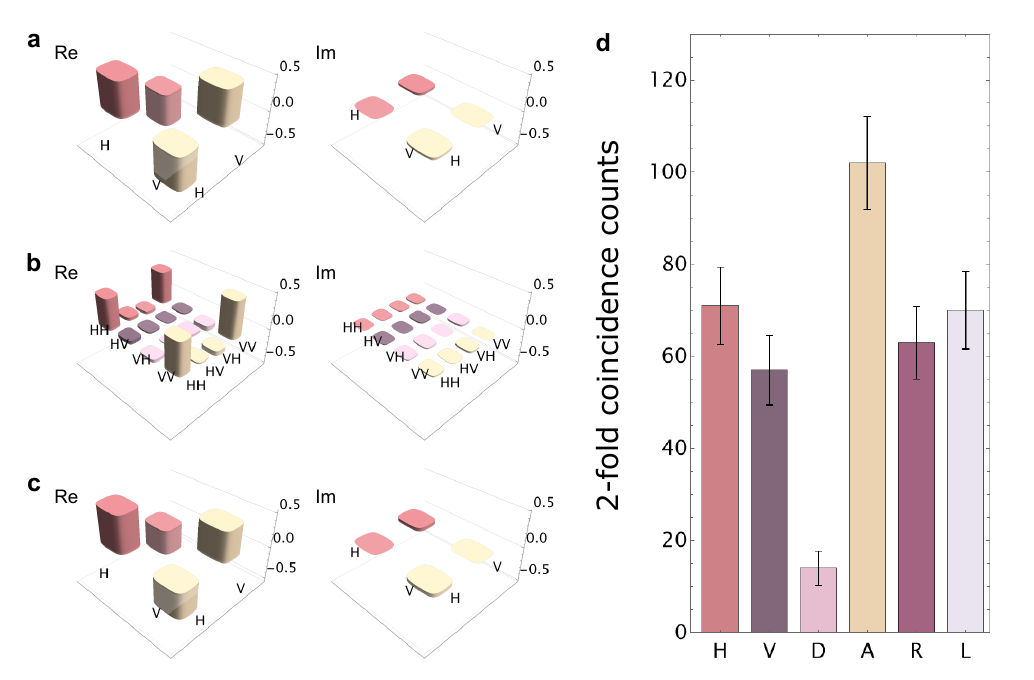}
  \caption{{\bf Experimental results for the SFG-based quantum teleportation.} {\bf a,}~Density matrix of 
  the $A$-polarized input state. The left~(right) of each plot shows the real~(imaginary) part of the density matrix, 
  respectively. {\bf b,}~Density matrix of the entangled state. {\bf c,}~Density matrix of the teleported state. 
{\bf d,}~Raw counts of the teleported photons for the $A$-polarized input light. The measurement time was 
  \SI{13}{h} for each basis state. Here, $R$ and $L$ represent right and left circular polarizations, respectively. The error bars were calculated assuming the Poisson statistics.  
  \label{fig3:Teleportation}} 
 \end{center}
\end{figure}

\begin{figure}[t]
 \begin{center}
  \includegraphics[width=\columnwidth]{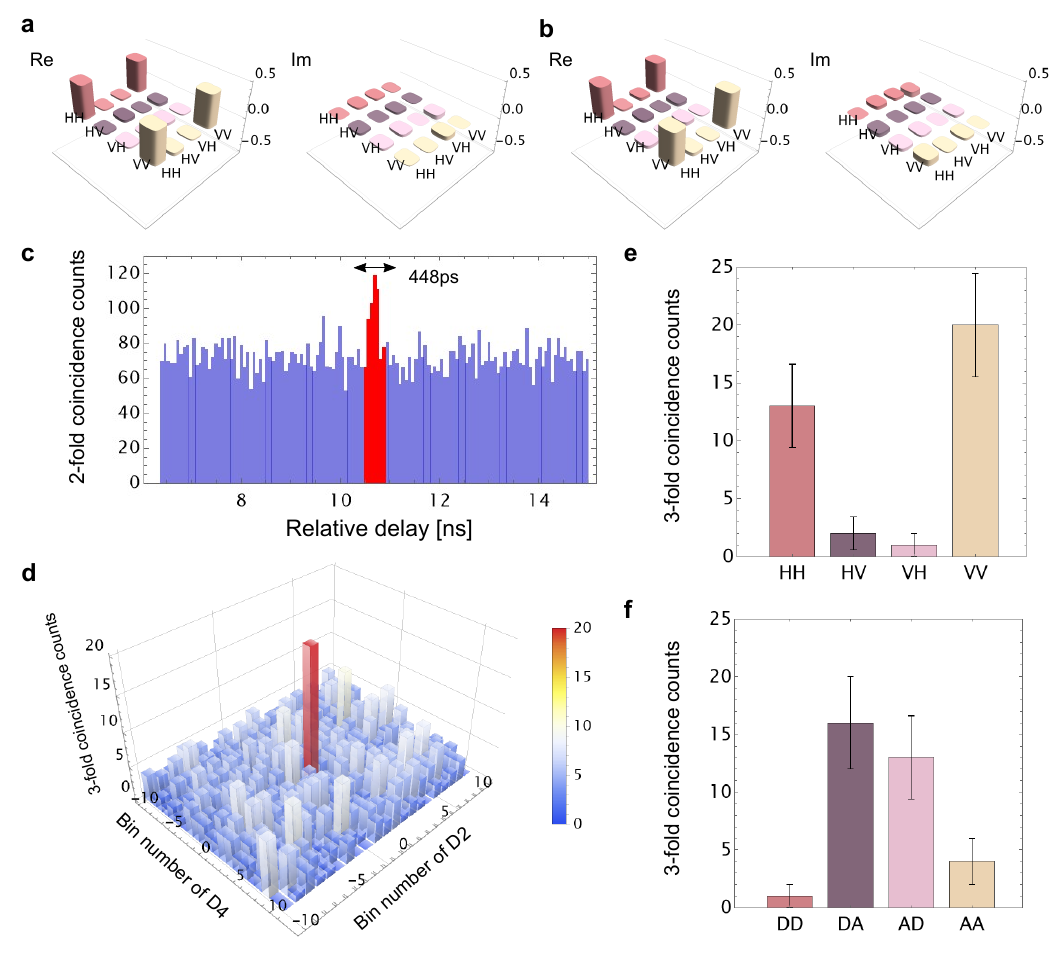}
  \caption{{\bf Experimental results for the SFG-based entanglement swapping.} 
  {\bf a, b,}~Density matrices of the input states from EPS~I and II, respectively.
  {\bf c,}~Two-fold coincidence counts between D1 and D2. We employed a 448-ps 
  coincidence window corresponding to seven bins around the signal peak. 
  {\bf d,}~Three-fold coincidence counts among D1, D2, and D4. The center bin 
  corresponds to the signal event. The waveplates were set so that D2 and D4 
  detect $V$-polarized portions of photons $d$ and $e$, respectively. 
  The measurement time was \SI{226}{h}. {\bf e, f,}~Raw detection counts 
  of the swapped state for each basis state. The error bars were 
  calculated assuming the Poisson statistics.
  \label{fig4:Swapping}} 
 \end{center}
\end{figure}

\clearpage

\renewcommand{\thefigure}{S\arabic{figure}}
\setcounter{table}{0}
\renewcommand{\thetable}{S\arabic{table}}
\setcounter{figure}{0}
\renewcommand{\theequation}{S\arabic{equation}}
\setcounter{equation}{0}

\section*{Supplementary information for \enquote{Experimental entanglement swapping
through single-photon $\chi^{(2)}$ nonlinearity}}

\section*{SUPPLEMENTARY NOTE 1: MODEL OF SFG-BASED ENTANGLEMENT SWAPPING}

We present a detailed procedure for calculating the theoretical visibilities of the swapped state using the realistic model shown in Supplementary Fig.~\ref{figS1:SFGmodel}. 
We assume that each EPS consists of two two-mode squeezed vacua, and the initial state is given by
$\ket{\psi_{in}}=\ket{\psi}_{ad}\otimes\ket{\psi}_{be}$, where 
\begin{equation}
\ket{\psi}_{ad}:=\sqrt{1-\gamma_{1H}^2}\sum_{k=0}^\infty\frac{1}{k!}(\gamma_{1H}\hat{a}^\dagger_{H}\hat{d}^\dagger_{H})^k\vac\otimes\sqrt{1-\gamma_{1V}^2}\sum_{l=0}^\infty\frac{1}{l!}(\gamma_{1V}\hat{a}^\dagger_{V}\hat{d}^\dagger_{V})^l\vac
\end{equation}
and
\begin{equation}
\ket{\psi}_{be}:=\sqrt{1-\gamma_{2H}^2}\sum_{k=0}^\infty\frac{1}{k!}(\gamma_{1H}\hat{b}^\dagger_{H}\hat{e}^\dagger_{H})^k\vac\otimes\sqrt{1-\gamma_{2V}^2}\sum_{l=0}^\infty\frac{1}{l!}(\gamma_{1V}\hat{b}^\dagger_{V}\hat{e}^\dagger_{V})^l\vac. 
\end{equation}
Here, the photon-number distributions are characterized by the average photon numbers as $\gamma_{1H}=\sqrt{\mu_{1H}/(1+\mu_{1H})}$ for example. 
Hereafter, we consider the events where up to a total of three photon pairs are produced in $\ket{\psi_{in}}$. 
Optical losses in mode $a_H$, $a_V$, $b_H$ and $b_V$ are simulated by virtual beamsplitters~(BSs) whose transmittance 
are $t_{1H}$, $t_{1V}$, $t_{2H}$ and $t_{2V}$, respectively. 
For example, the optical loss operation in mode $a_{H}$ is represented by 
\begin{equation}
\mathcal{L}_{a'_H}(\ketbra{\psi_{in}}{\psi_{in}})=\tr_{a'_{H}}[\hat{U}_{a'_H}\ketbra{\psi_{in}}{\psi_{in}}\otimes\ket{\mathrm{vac}}_{a'_{H}}\bra{\mathrm{vac}}_{a'_{H}}\hat{U}_{a'_H}^\dagger], 
\end{equation}
where $a'_H$ is an ancillary mode and $\hat{U}_{a'_H}$ is the unitary operator of the BS whose transmittance is $t_{1H}$, which satisfies
$\hat{U}_{a'_H}\hat{a}^\dagger_{H}\hat{U}_{a'_H}^\dagger=\sqrt{t_{1H}}\hat{a}^\dagger_{H}+\sqrt{1-t_{1H}}\hat{a'}^\dagger_{H}$ and $\hat{U}_{a'_H}\vac=\vac$. 
The quantum state after experiencing the optical losses is given by 
\begin{equation}
\hat{\rho}\u{L}=\mathcal{L}_{a'_H}\circ\mathcal{L}_{a'_V}\circ\mathcal{L}_{b'_H}\circ\mathcal{L}_{b'_V}(\ketbra{\psi_{in}}{\psi_{in}}). 
\label{eqS1:rhoL}
\end{equation}

\begin{figure}[t]
 \begin{center}
  \includegraphics[width=\columnwidth]{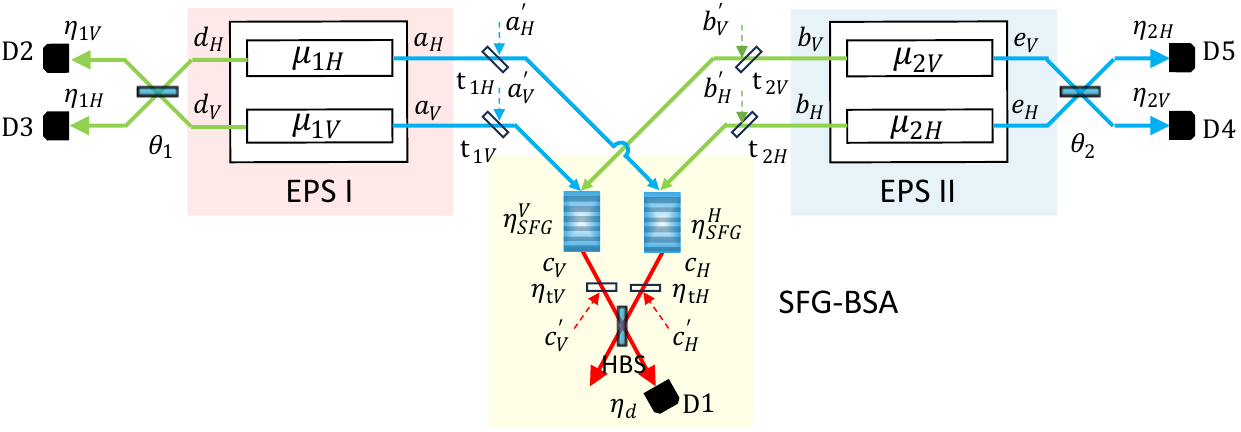}
  \caption{{\bf Realistic model of SFG-based entanglement swapping.} Each EPS consists of two photon pair sources. The polarization correlation measurements on the swapped state are performed by the beamsplitters in the polarization DOF followed by photon detections using threshold detectors. 
  \label{figS1:SFGmodel}} 
 \end{center}
\end{figure}

Since the threshold detectors are used in the later photon detection, we only need to consider the subspace of $\hat{\rho}\u{L}$ where the number of photons in $\ket{\psi_{in}}$ and $\bra{\psi_{in}}$ are equal.
As a next step, the SFG operation is performed on the photons in mode $a_H$, $a_V$, $b_H$ and $b_V$. 
We note that Eq.~(\ref{eq2:QPC}) in the main text is not enough to express the SFG operation in this model, since it is
derived under the condition that the number of input photons is at most two. 
Thus, we need to derive the SFG operation on the three-photon inputs. 
Moreover, we consider that the coupling constant $\chi$ in Eq.~(\ref{eq1:SFG}) in the main text depends on the polarization of the
inputs, which we define $\chi_{H(V)}$ for $H(V)$-polarized input, respectively. 
Considering only the first order of $\chi_{H}$ and $\chi_{V}$ in $U=e^{-iH'\tau/\hbar}$, the corresponding operation is approximated by 
\begin{equation}
U\sim-i\hat{H}'\tau/\hbar=\tau(\chi_H\hat{a}_H\hat{b}_H\hat{c}^\dagger_H+\chi_V\hat{a}_V\hat{b}_V\hat{c}^\dagger_V)-\mathrm{H.c.}.
\label{eqS2:SFG}
\end{equation}
We exemplify the dynamics for the SFG operations involving three photons. 
As the first case, we consider the case where two photons $\hat{a}^\dagger_i$ and $\hat{b}^\dagger_i$ in an identical polarization state and a photon $\hat{a}^\dagger_j$ in a polarization orthogonal to them are input, e.g. $\hat{a}^{\dagger}_H\hat{a}^{\dagger}_V\hat{b}^\dagger_H\vac$. 
Performing Supplementary Eq.~(\ref{eqS2:SFG}), we obtain 
\begin{equation}
(-i\hat{H}'\tau/\hbar)\hat{a}^{\dagger}_H\hat{a}^{\dagger}_V\hat{b}^\dagger_H\vac=\sqrt{\eta\u{SFG}^H}\hat{a}^{\dagger}_V\hat{c}^\dagger_H\vac, 
\end{equation}
where $\eta\u{SFG}^H:=(\chi_H\tau)^2$.
As the second case, we consider the case where two photons $(\hat{a}^\dagger_i)^2$ in an identical polarization state
and photon $\hat{b}^\dagger_j$ in a polarization orthogonal to that of a photon $a$ are input. For example, $(\hat{a}^{\dagger}_H)^2\hat{b}^\dagger_V\vac$ is transformed into 
\begin{equation}
(-i\hat{H}'\tau/\hbar)(\hat{a}^{\dagger}_H)^2\hat{b}^\dagger_V\vac=0.  
\end{equation}
As the third case, we consider the case where the three input photons are in an identical polarization state. For example, $\frac{1}{\sqrt{2}}(\hat{a}^{\dagger}_H)^2\hat{b}^\dagger_H\vac$ is transformed into  
\begin{eqnarray}
(-i\hat{H}'\tau/\hbar)\frac{1}{\sqrt{2}}(\hat{a}^{\dagger}_H)^2\hat{b}^\dagger_H\vac
&=&\sqrt{\frac{\eta\u{SFG}^H}{2}}(1+\hat{a}^\dagger_H\hat{a}_H)\hat{a}^{\dagger}_H(1+\hat{b}^\dagger_H\hat{b}_H)\hat{c}^\dagger_H\vac \\
&=&\sqrt{2\eta\u{SFG}^H}\hat{a}^{\dagger}_H\hat{c}^\dagger_H\vac, 
\end{eqnarray}
which shows that the SFG efficiency is doubled compared to the case where $\hat{a}^{\dagger}_H\hat{a}^{\dagger}_V\hat{b}^\dagger_H\vac$ is the input.
The SFG photons in mode $c_H$ and $c_V$ experience optical losses $\mathcal{L}_{c'_{H}}$ and $\mathcal{L}_{c'_{V}}$, respectively. 
The projection of the SFG photon on $\ket{D}$ and $\ket{A}$ is simulated by mixing the photons in mode $c_H$ and $c_V$ by a virtual HBS in polarization degree of freedom
followed by photon detections by threshold detectors.
Since we consider the case where at most one SFG photon is emitted, 
POVM elements of the projective measurements on $\ket{D}$ and $\ket{A}$ are approximated by
$\Pi^{D}_{c}\sim\eta_d\ket{D}_c\bra{D}_c$ and $\Pi^{A}_{c}\sim\eta_d\ket{A}_c\bra{A}_c$,
respectively, where $\eta_d$ is the quantum efficiency of D1. 
The unnormalized quantum state after detecting the $A$-polarized SFG photon is given by 
\begin{equation}
\hat{\rho}\u{SFG}=\tr_{a_H,a_V,b_H,b_V,c_H,c_V}[\Pi^A_{c}\mathcal{L}_{c'_{H}}\circ\mathcal{L}_{c'_{V}}((-i\hat{H}'\tau/\hbar)\hat{\rho}\u{L}(-i\hat{H}'\tau/\hbar)^\dagger)]. 
\label{eqS3:rhoSFG}
\end{equation}
Finally, polarization correlation measurements on the photons in mode $d$ and $e$ are performed. 
Similarly to the projective measurement on the SFG photon, the $H$- and $V$-polarized components are mixed by virtual BSs in polarization degree of freedom whose transmittance is $\cos^2\theta_1$ and $\cos^2\theta_2$ for the photons in mode $d$ and $e$, respectively, and are
detected by the threshold detectors. 
We exemplify the POVM element of photon detection at D3 as 
\begin{equation}
\hat{\Pi}_{d_H}(\theta_1)=\sum_{n=1}^2\frac{1}{n!}(1-(1-\eta_{1H})^n)\hat{U}\u{BS}(\theta_1)(\hat{d}^\dagger_{H})^n\ketbra{\mathrm{vac}}{\mathrm{vac}}(\hat{d}_{H})^n\hat{U}\u{BS}^\dagger(\theta_1). 
\label{eqS3:POVM}
\end{equation}
Here, the sum of $n$ is limited to 2, since the maximum number of photons in a single mode is 2 in this model.
$\hat{U}\u{BS}(\theta_1)$ is the unitary operator of the BS, which satisfies
$\hat{U}\u{BS}(\theta_1)\hat{d}^\dagger_{H}\hat{U}\u{BS}^\dagger(\theta_1)=\cos\theta_1\hat{d}^\dagger_{H}+\sin\theta_1\hat{d}^\dagger_{V}$ and $\hat{U}\u{BS}(\theta_1)\vac=\vac$. When Z(X)-basis measurement is performed, $\theta_1$ and $\theta_2$ are set to be 0($\pi/4$), respectively. 
For example, the coincidence probability between D3 and D5 heralded by D1 is given by 
\begin{equation}
P^\mathrm{SFG}_{HH}(\theta_1,\theta_2)=\tr[\hat{\Pi}_{d_H}(\theta_1)\hat{\Pi}_{e_H}(\theta_2)\hat{\rho}\u{SFG}].  
\label{eqS4:Ps}
\end{equation}
Here the subscript \enquote{$HH$} comes from the fact that D3 and D5 detect 
$H$-polarized components after the FPBSs.~(See Fig.~\ref{fig2:Experiment} in the main text)

In addition to the influence of the multiphoton creation considered above,
there are accidental coincidences caused by the dark counts of the detectors. 
Since the dark count probabilities of D2, D3, D4, and D5 are sufficiently smaller than their photon detection probabilities, only the effect of the dark count in D1 is considered in this model.
The coincidence probability between D3 and D5 heralded by the dark count in D1 is given by 
$R_d\tau_W(P^\mathrm{Acd}_{HH}(\theta_1,\theta_2)-P^\mathrm{SFG}_{HH}(\theta_1,\theta_2))$, where 
$R_d\tau_W$ is the dark count probability of D1, and 
\begin{equation}
P^\mathrm{Acd}_{HH}(\theta_1,\theta_2)=\tr[\hat{\Pi}_{d_H}(\theta_1)\hat{\Pi}_{e_H}(\theta_2)\tr_{aH,aV,bH,bV}[\ketbra{\psi_{in}}{\psi_{in}}]] 
\label{eqS5:Pd}
\end{equation}
is the accidental coincidence between D3 and D5. 
\begin{table}[h]
\caption{{\bf The experimental parameters necessary for calculating $V\u{Z}^\mathrm{th}$ and $V\u{X}^\mathrm{th}$.}}\label{tableS1:Parameters}%
\begin{tabular}{@{}llllllllll@{}}
\toprule
$\ast$ &  $\mu_{1*}$ & $\eta_{1*}$ & $t_{1*}$&  $\mu_{2*}$ & $\eta_{2*}$ & $t_{2*}$ & $\eta^*\u{SFG}$ & $\eta_{t*}$ & $\eta_d$\\
\midrule    
$H$&  0.060 & 0.097 & 0.44 &  0.080 & 0.070 & 0.56 & $2.31\times10^{-8}$ & 0.43 & 0.85\\
$V$&  0.050 & 0.11 & 0.48 &  0.061 & 0.10 & 0.57 & $2.35\times10^{-8}$ & 0.40 & 0.85\\
\botrule
\end{tabular}
\end{table}
By using Supplementary Eq.~(\ref{eqS4:Ps}) and Supplementary Eq.~(\ref{eqS5:Pd}), the visibilities in Z and X basis are respectively given by 
\begin{equation}
V\u{Z}^\mathrm{th}=\frac{P_{HH}(0,0)+P_{VV}(0,0)-P_{HV}(0,0)-P_{VH}(0,0)}{P_{HH}(0,0)+P_{VV}(0,0)+P_{HV}(0,0)+P_{VH}(0,0)}
\label{eqS6:VZ}
\end{equation}
and
\begin{equation}
V\u{X}^\mathrm{th}=\frac{P_{HV}(\frac{\pi}{4},\frac{\pi}{4})+P_{VH}(\frac{\pi}{4},\frac{\pi}{4})-P_{HH}(\frac{\pi}{4},\frac{\pi}{4})-P_{VV}(\frac{\pi}{4},\frac{\pi}{4})}{P_{HV}(\frac{\pi}{4},\frac{\pi}{4})+P_{VH}(\frac{\pi}{4},\frac{\pi}{4})+P_{HH}(\frac{\pi}{4},\frac{\pi}{4})+P_{VV}(\frac{\pi}{4},\frac{\pi}{4})}, 
\label{eqS7:VX}
\end{equation}
where $P_{ij}(\theta_1,\theta_2):=P^\mathrm{SFG}_{ij}(\theta_1,\theta_2)(1-R_d\tau_W)+R_d\tau_WP^\mathrm{Acd}_{ij}(\theta_1,\theta_2)$ with $i,j\in\{H, V\}$. 
Here, $V\u{Z}^\mathrm{th}$ and $V\u{X}^\mathrm{th}$ are functions of the experimental parameters summarized in Supplementary Table~\ref{tableS1:Parameters}.
Substituting the experimental parameters, we obtain $V\u{Z}^\mathrm{th}=0.78$ and $V\u{X}^\mathrm{th}=0.76$, respectively.

\section*{SUPPLEMENTARY NOTE 2: COMPARISON OF SFG-BSA AND LINEAR OPTICAL BSA}
\begin{figure}[t]
 \begin{center}
  \includegraphics[width=\columnwidth]{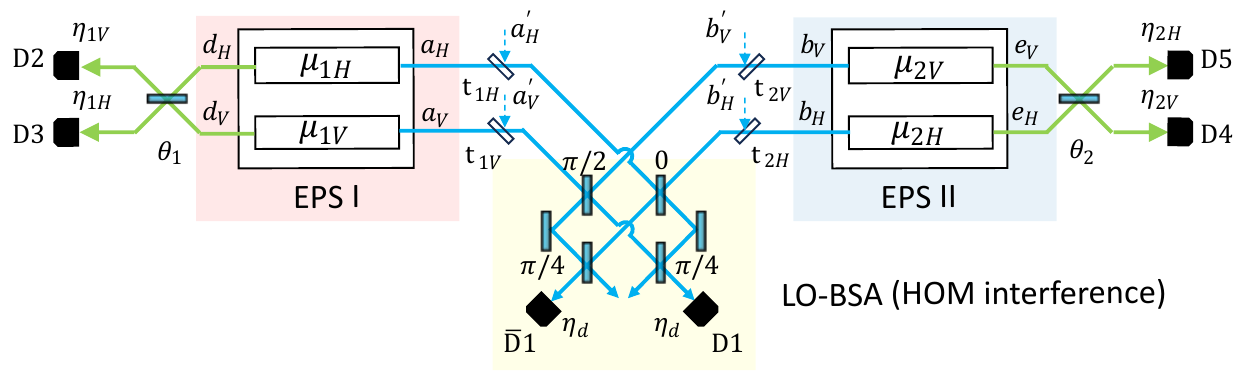}
  \caption{{\bf Model of linear optical entanglement swapping.} 
Linear optical BSA is realized by mixing the photons in mode $a$ and $b$ with a polarization beamsplitter and then performing projective measurements in X basis on each output.
  \label{figS2:LOmodel}} 
 \end{center}
\end{figure}

In this section, we quantitatively show the superiority of the SFG-BSA over the linear optical BSA~(LO-BSA). We introduce the model of the LO-BSA as shown in Supplementary Fig.~\ref{figS2:LOmodel}. 
We assume perfect HOM interference and no dark count at the BSA. 
The unnormalized quantum state just before the BSA is given by $\hat{\rho}\u{L}$ in Supplementary Eq.~(\ref{eqS1:rhoL}). 
Here, we assume the symmetric transmittance as $t_{1H}=t_{1V}=t_{2H}=t_{2V}=t$ for simplicity. 
The photons in mode $a$ and $b$ are then mixed by a PBS, which is simulated by 
mixing the $H$- and $V$-polarized components of the photons using two beamsplitters 
$\hat{U}^H\u{BS}(0)$ and $\hat{U}^V\u{BS}(\pi/2)$, respectively as
\begin{equation}
\hat{\rho}\u{PBS}=(\hat{U}^H\u{BS}(0)\otimes\hat{U}^V\u{BS}(\pi/2))\hat{\rho}_L(\hat{U}^H\u{BS}(0)\otimes\hat{U}^V\u{BS}(\pi/2))^\dagger, 
\end{equation}
where $\hat{U}^H\u{BS}(0)$ and $\hat{U}^V\u{BS}(\pi/2)$ satisfy $\hat{U}^H\u{BS}(0)\hat{a}_H^\dagger(\hat{U}^H\u{BS}(0))^\dagger=\hat{a}^\dagger_H$, $\hat{U}^H\u{BS}(0)\hat{b}_H^\dagger(\hat{U}^H\u{BS}(0))^\dagger=\hat{b}^\dagger_H$, 
$\hat{U}^V\u{BS}(\pi/2)\hat{a}_V^\dagger(\hat{U}^V\u{BS}(\pi/2))^\dagger=\hat{b}^\dagger_V$, $\hat{U}^V\u{BS}(\pi/2)\hat{b}_V^\dagger(\hat{U}^V\u{BS}(\pi/2))^\dagger=\hat{a}^\dagger_V$ and $\hat{U}^H\u{BS}(0)\vac=\hat{U}^V\u{BS}(\pi/2)\vac=\vac$. 
Finally, the four-fold coincidences among D1, $\mathrm{\bar{D}1}$, (D2 or D3) and (D4 or D5) are measured. 
Here, $A$- and $D$-polarized components are respectively detected by D1 and $\mathrm{\bar{D}}1$ whose quantum efficiencies are $\eta_d$. 
For example, the four-fold coincidence probability among D1, $\mathrm{\bar{D}}1$, D3 and D5 is given by using the POVM elements defined in Supplementary Eq.~(\ref{eqS3:POVM}) as 
\begin{equation}
P^\mathrm{LO}_{HH}(\theta_1,\theta_2)=\tr[\hat{\Pi}_{dH}(\theta_1)\hat{\Pi}_{eH}(\theta_2)\hat{\Pi}_{b_H}(\pi/4)\hat{\Pi}_{a_V}(\pi/4)\hat{\rho}\u{PBS}]. 
\label{eqS9:rhoH}
\end{equation}
The visibilities of the swapped state are calculated by replacing $P_{ij}(\theta_1,\theta_2)$ in Eqs.~(\ref{eqS6:VZ}) and (\ref{eqS7:VX}) with
$P^\mathrm{LO}_{ij}(\theta_1,\theta_2)$.

\begin{figure}[t]
 \begin{center}
  \includegraphics[width=\columnwidth]{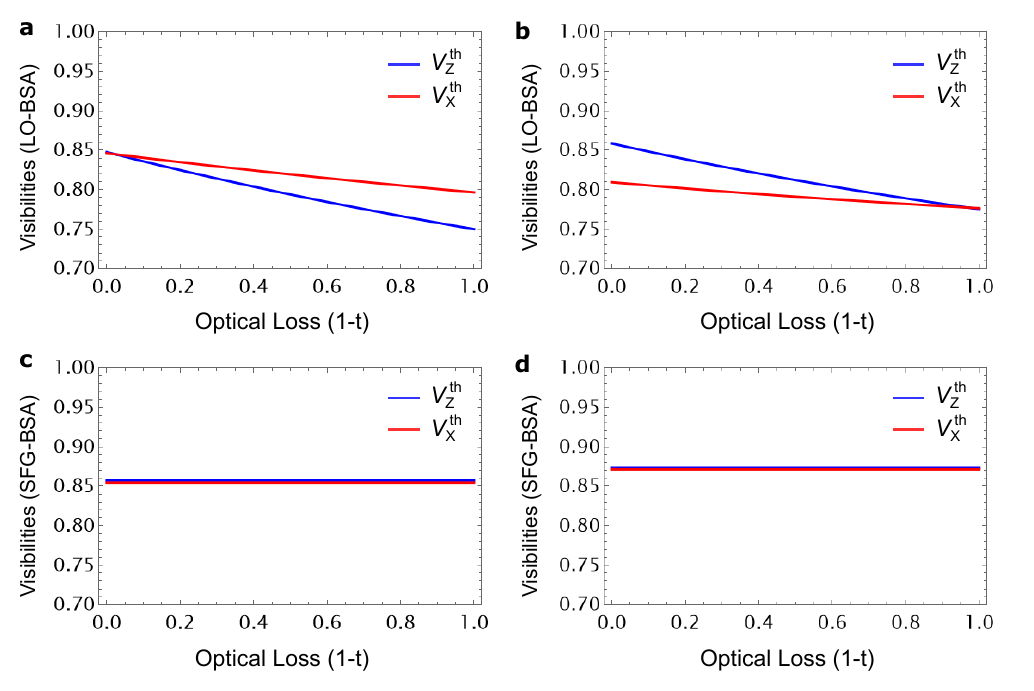}
  \caption{{\bf Visibilities versus optical loss.} {\bf a, b} Entanglement swapping using LO-BSA with {\bf a}~$\eta_{1H}=\eta_{1V}=\eta_{2H}=\eta_{2V}=1$ and {\bf b}~$\eta_{1H}=\eta_{1V}=\eta_{2H}=\eta_{2V}$=0.1. {\bf c,d} Entanglement swapping using SFG-BSA with {\bf c}~$\eta_{1H}=\eta_{1V}=\eta_{2H}=\eta_{2V}=1$ and {\bf d}~$\eta_{1H}=\eta_{1V}=\eta_{2H}=\eta_{2V}=0.1$. 
  \label{figS3:LOvsSFG}} 
 \end{center}
\end{figure}

In the following, we assume that the average photon number of each two-mode squeezed vacuum is the same as $\mu_{1H}=\mu_{1V}=\mu_{2H}=\mu_{2V}=0.05$ 
and the photon detection efficiencies at the LO-BSA are unity as $\eta_d=1$ for simplicity. 
The visibilities $V^\mathrm{th}\u{Z}$ and $V^\mathrm{th}\u{X}$ versus optical loss $(1-t)$ 
with $\eta_{1H}=\eta_{1V}=\eta_{2H}=\eta_{2V}=1$ and $\eta_{1H}=\eta_{1V}=\eta_{2H}=\eta_{2V}=0.1$
are plotted in Supplementary Fig.~\ref{figS3:LOvsSFG}{\bf a} and {\bf b}, respectively. 
The visibilities decrease as the photon losses increase. 
This is because the portion of fake success events caused by two photons from one EPS to enter the LO-BSA and be a coincidence between D1 and $\mathrm{\bar{D}}1$ increases as the photon losses increase. 

The visibilities obtained by the SFG-BSA {\it without} dark counts are calculated
by replacing $P_{ij}(\theta_1,\theta_2)$ in Eqs.~(\ref{eqS6:VZ}) and (\ref{eqS7:VX}) with
$P^\mathrm{SFG}_{ij}(\theta_1,\theta_2)$. 
Here, we assume $\mu_{1H}=\mu_{1V}=\mu_{2H}=\mu_{2V}=0.05$, $t_{1H}=t_{1V}=t_{2H}=t_{2V}=t$ and
$\eta_{TH}=\eta_{TV}=\eta_d=1$ as in the case of the LO-BSA. 
The visibilities $V^\mathrm{th}\u{Z}$ and $V^\mathrm{th}\u{X}$ versus optical loss $(1-t)$ with $\eta_{1H}=\eta_{1V}=\eta_{2H}=\eta_{2V}=1$ and $\eta_{1H}=\eta_{1V}=\eta_{2H}=\eta_{2V}=0.1$ are plotted
in Supplementary Fig.~\ref{figS3:LOvsSFG}{\bf b} and {\bf d}, respectively. 
We see that the visibilities are insensitive to the optical losses. 
This is because the fake success events related to the optical losses, which occur in the LO-BSA, are rejected by the SFG-BSA. 
As a result, the visibilities remain high despite the effects of optical losses.

\section*{SUPPLEMENTARY NOTE 3: EXPERIMENTAL RESULTS OF SFG-BASED QUANTUM TELEPORTATION}
We show all density matrices and raw counts obtained in the quantum teleportation experiment using SFG-BSA. 
The density matrices of the input states are shown in Supplementary Fig.~\ref{figS4:inputs}. 
The density matrices and raw counts of the teleported states are shown in Supplementary Fig.~\ref{figS5:final_T} {\bf a}-{\bf c} and {\bf d}-{\bf f}, respectively.

\begin{figure}[t]
 \begin{center}
  \includegraphics[width=\columnwidth]{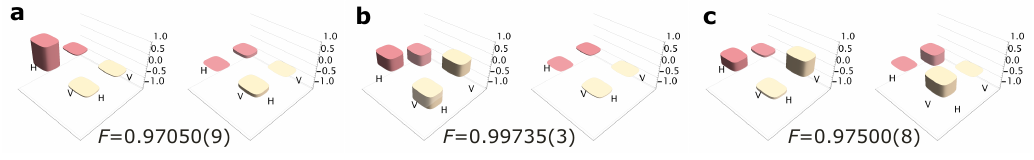}
  \caption{{\bf Input polarization states.} {\bf a}, {\bf b}, {\bf c} Density matrices of the $H$-, $A$-, and $R$-polarized states, respectively. 
  \label{figS4:inputs}} 
 \end{center}
\end{figure}

\begin{figure}[t]
 \begin{center}
  \includegraphics[width=\columnwidth]{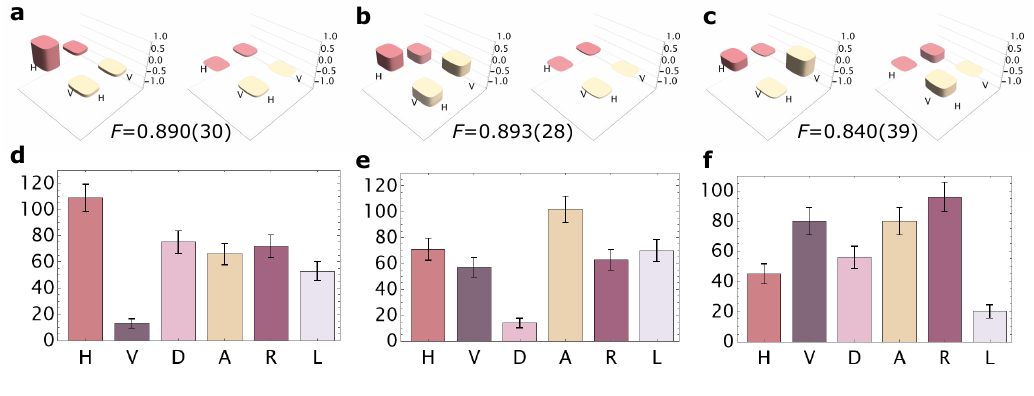}
  \caption{{\bf Teleported states and raw detection counts.} {\bf a}, {\bf b}, {\bf c} Density matrices of the teleported states for the $H$-, $A$-, and $R$-polarized input states, respectively. {\bf d}, {\bf e}, {\bf f} Raw detection counts of the teleported photons for the $H$-, $A$-, and $R$-polarized input states, respectively. The error bars were calculated assuming the Poisson statistics.
  \label{figS5:final_T}} 
 \end{center}
\end{figure}

\section*{SUPPLEMENTARY NOTE 4: SFG-BASED QUANTUM TELEPORTATION WITH LARGE-PHOTON-NUMBER INPUTS}

\begin{figure}[t]
 \begin{center}
  \includegraphics[width=\columnwidth]{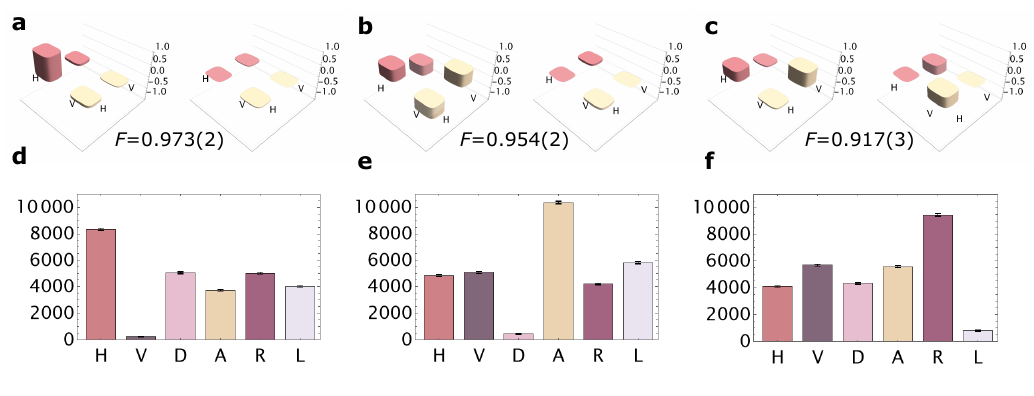}
  \caption{{\bf Output states and raw detection counts.} {\bf a}, {\bf b}, {\bf c} Density matrices of the output states after the QFC for the $H$-, $A$-, and $R$-polarized input states, respectively. {\bf d}, {\bf e}, {\bf f} Raw detection counts of the output photons after the QFC for the $H$-, $A$-, and $R$-polarized input states, respectively. The error bars were calculated assuming the Poisson statistics.
  \label{figS5:final_Q}} 
 \end{center}
\end{figure}

We used coherent light with an average photon number of $7.5\times10^4$ to prepare the input states and perform the quantum teleportation experiment. 
We show the density matrices and raw counts of the output states in Supplementary Fig.~\ref{figS5:final_Q}. The fidelities of the output states for $H$-, $A$- and $R$-polarized inputs are $0.973(2)$, $0.954(2)$ and $0.917(3)$, respectively. 
Interestingly, unlike the case of the linear optical BSA, 
we observe the \enquote{quantum-teleportation-like} operation even if a number of multiple photons are contained in the input coherent light as reported in Ref.~\cite{PhysRevLett.86.1370}. 
This result can be interpreted as follows. 
Assuming that the average intensities of the $H$- and $V$-polarized portion of the input coherent light in mode $b$ are respectively given by $|\alpha|^2$ and $|\beta|^2$, where the polarization is expressed as $(\alpha {\textit {\textbf H}}+ \beta {\textit {\textbf V}})/\sqrt{|\alpha|^2+|\beta|^2}$, and are sufficiently large,
$\hat{b}_{H(V)}$ in Eq.~(1) in the main text is replaced with $\alpha(\beta)\in\mathbb{C}$, respectively, as  

\begin{equation}
\hat{H}\u{QFC}=i\hbar\chi(\alpha\hat{a}_H\hat{c}^\dagger_H+\beta\hat{a}_V\hat{c}^\dagger_V)+\mathrm{H.c.}, 
\label{eq7:QFC}
\end{equation}
which corresponds to the Hamiltonian of the quantum frequency conversion~(QFC)~\cite{Kumar:90,Tanzilli2005,Ikuta2011}, between mode $a$ and $c$. 
Here, the input light in mode $b$ is regarded as pump light. 
The unitary transformation $\hat{U}\u{QFC}=e^{-i\hat{H}\u{QFC}\tau/\hbar}$ of $\hat{a}_{H(V)}$ is represented by~\cite{Ikuta2018}
\begin{equation}
\hat{U}\u{QFC}\hat{a}_{H(V)}\hat{U}^\dagger\u{QFC}= \cos{(|\alpha(\beta)|\chi\tau)}\hat{a}_{H(V)}+e^{i\mathrm{arg}\alpha(\beta)}\sin{(|\alpha(\beta)|\chi\tau)}\hat{c}_{H(V)}.
\label{eq8:QPCK}
\end{equation}
In the weak pump regime where $|\alpha(\beta)|\chi\tau\ll1$, the right hand side of Supplementary Eq.~(\ref{eq8:QPCK}) is approximated by
$\hat{a}_{H(V)}+\sqrt{\eta\u{SFG}}\alpha(\beta)\hat{c}_{H(V)}$. 
Thus, after the QFC followed by the detection of a $D$-polarized SFG photon in mode $c$, we obtain
$_c\bra{D}\hat{U}\u{QFC}\ket{\Phi^+}_{ad}\propto\alpha\ket{H}_d+\beta\ket{V}_d$,
which shows that the polarization information of the input coherent light is transferred to the photon in mode $d$. 
We emphasize that the above dynamics is different from the quantum teleportation. 
Therefore, in the strong pump regime, the conversion efficiency is no longer proportional to the pump intensity, and the
situation is rather similar to the polarization insensitive QFC~\cite{Ikuta2018, Bock2018, PhysRevLett.124.010510} where no polarization information of the input~(pump) light remains in the output photon. 

\section*{SUPPLEMENTARY NOTE 5: FABRICATION OF PPLN/W3}
A ridge waveguide was formed by mechanical processing using a 3-inch z-cut MgO-doped PPLN substrate with a period of \SI{18.2}{\mu m}. The PPLN substrate was bonded to an undoped $\mathrm{LiNbO_3}$ substrate with an adhesive layer as an under-cladding. The PPLN substrate was thinned to a thickness of \SI{7.2}{\mu m} by a lapping and polishing process. The variation in PPLN thickness along the waveguide was within approximately \SI{0.5}{\mu m}. In the next step, the ridge structure was formed using a dicing saw. The ridge width was \SI{8}{\mu m}. To minimize chipping and waveguide width variation, the finest diamond blade was used to cut the PPLN film at low cutting speed, resulting in low propagation loss. The waveguide was cut to an interaction length of \SI{6.3}{cm}. The end faces of the waveguide device were anti-reflective coated for both telecom and \SI{780}{nm} light.

\section*{SUPPLEMENTARY NOTE 6: SPATIAL PROFILES AND SPECTRAL DISTRIBUTIONS}
The spatial profiles and spectral distributions of the photons that are involved in the 
single-photon SFG process are shown in Supplementary Fig.~\ref{figS0:Modeprofile}{\bf a} and {\bf b}, respectively. The spatial profiles of the photons in mode $a$ and $b$ were simulated by injecting laser light at \SI{1535}{nm} and \SI{1585}{nm} into PPLN/W1 and 2, respectively. The spatial profile of the photon in mode $c$ was simulated by measuring the spatial profile of the SFG light at \SI{780}{nm} generated by PPLN/W3. The spectral distributions of the photons in mode $a$ and $b$ were estimated by measuring the diffraction spectrum of the VHG and transmission spectrum of Filter~I by sweeping the center wavelength of the tunable laser, respectively. 
The spectral distribution of the photon in mode $c$ was estimated by measuring the phase matching bandwidth of PPLN/W3. 
The FWHMs of the bandwidths of the photons in mode $a$ and $c$ were estimated to be \SI{55}{GHz} and \SI{39}{GHz}, respectively. 
As shown in Supplementary Fig.~\ref{figS0:Modeprofile}{\bf b}, the bandwidth of the photon in mode $b$ was much broader than those of the photons in mode $a$ and $c$. 
In fact, a narrow band pass filter was not necessary for the photon in mode $b$ in our setup because the components that were
outside the phase matching bandwidth of PPLN/W3 and bandwidth of the photon in mode $a$ do not contribute to the SFG process. 
\begin{figure}[t]
 \begin{center}
  \includegraphics[width=\columnwidth]{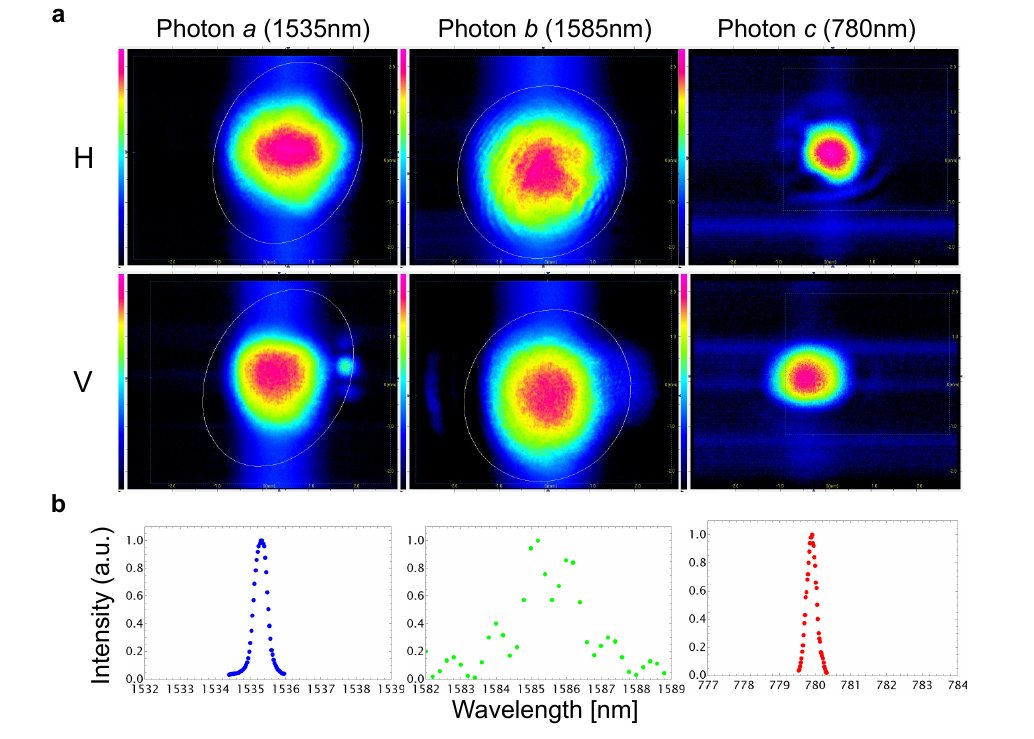}
  \caption{{\bf Spatial profiles and spectral distributions.} Spatial profiles ({\bf a}) and
  spectral distributions ({\bf b}) of the photons in mode $a$, $b$ and $c$. 
  \label{figS0:Modeprofile}} 
 \end{center}
\end{figure}

\section*{SUPPLEMENTARY NOTE 7: APPLICATION TO LOOPHOLE-FREE BELL TEST}
We describe here the details of the simulation used to evaluate the performance of SFG-BSA for a loophole-free Bell test and device-independent quantum key distribution~(DIQKD). We begin by outlining the procedure for performing the Bell test using the quantum state heralded by SFG-BSA. 
As illustrated in Supplementary Fig.~\ref{figS1:SFGmodel}, under the condition that D1 clicks, Alice and Bob independently choose their measurement angles $ \{\theta_{A1}, \theta_{A2}\}\in\theta_1$ and $\{\theta_{B1}, \theta_{B2}\}\in\theta_2$, respectively, and perform measurements. For each setting, they record all combinations of detection (\enquote{click}) and non-detection (\enquote{no-click}) events without any postselection. 
Each party then applies a locally defined strategy to assign binary outcomes ($\pm1$) to each detection event. Since each party has four possible local outcomes--(i) only D3 (D5) clicks, (ii) only D2 (D4) clicks, (iii) both detectors click, and (iv) neither detector clicks--there are $2^4 = 16$ possible strategies per party.
In our simulation, we adopt the following representative strategy: outcome (i) is assigned $-1$, while outcomes (ii), (iii), and (iv) are assigned $+1$. For example, when Alice and Bob choose measurement settings $\theta_{A1}$ and $\theta_{B1}$, respectively, the probability that both obtain the outcome $-1$ which 
we denote $P(-1,-1|\theta_{A1},\theta_{B1})$ is obtained by the probability that detectors D3 and D5 click, while detectors D2 and D4 do not. Similarly, the conditional probabilities $P(+1,-1|\theta_{A1},\theta_{B1})$, $P(-1,+1|\theta_{A1},\theta_{B1})$, and $P(+1,+1|\theta_{A1},\theta_{B1})$ are also computed from the corresponding detection events.
The Clauser-Horne-Shimony-Holt~(CHSH) value $S$ is defined by 
\begin{equation}
S=\expect{A_1B_1}+\expect{A_2B_1}+\expect{A_1B_2}-\expect{A_2B_2},
\label{eqS5:CHSH}
\end{equation}
where $\expect{A_iB_j}=P(+1,+1|\theta_{Ai},\theta_{Bj})+P(-1,-1|\theta_{Ai},\theta_{Bj})-P(+1,-1|\theta_{Ai},\theta_{Bj})-P(-1,+1|\theta_{Ai},\theta_{Bj})$.  
Although the maximal value of $|S|$ is in the upper bound of 2 in the framework of a local realism theory, it can exceed 2 in quantum mechanics. 
When Alice and Bob perform DIQKD, Alice chooses another measurement angle $\theta_{A0}$, and
the raw key is generated under the condition that Alice and Bob choose $\theta_{A0}$ and $\theta_{B1}$, respectively. 
The lower bound of the asymptotic key rate $r$ is represented by~\cite{PhysRevLett.98.230501,pironio2009device}
\begin{equation}
r\geq r_{\mathrm{DW}}=1-h(Q)-\chi(S), 
\label{eq:keyrate}
\end{equation}
where $r_{\mathrm{DW}}$ is the Devetak-Winter rate~\cite{devetak2005distillation}, $Q$ is qubit error rate which is defined by
\begin{equation}
Q=P(+1,-1|\theta_{A0},\theta_{B1})+P(-1,+1|\theta_{A0},\theta_{B1}),  
\label{eq:Qber}
\end{equation}
and 
\begin{equation}
\chi(S)=h\left[\frac{1+\sqrt{(S/2)^2-1}}{2}\right].  
\end{equation}
Here, $h(\cdot)$ is the binary entropy defined by $h(x)=-x\mathrm{log}_2x-(1-x)\mathrm{log}_2(1-x)$.

Next, we calculate the maximum CHSH value $S$ achieved by SFG-BSA. 
We consider the case where each detector has no dark count and has unity
detection efficiency, and the quantum state used for the Bell test is given by $\hat{\rho}\u{SFG}/\tr[\hat{\rho}\u{SFG}]$ in Supplementary Eq.~(\ref{eqS3:rhoSFG}). 
By optimizing the average photon number of each SPDC source and the measurement angles, 
we obtain the maximum CHSH value of $S=2.82843\sim2\sqrt{2}$, which is
the maximum value allowed by quantum mechanics known as the Tsirelson bound~\cite{Cirelson1980-ly} and
much higher than the maximum CHSH value achieved by the LO-BSA~\cite{Tsujimoto_2020}. 
By maximizing the CHSH value with decreasing $\eta_{1H}$, $\eta_{1V}$, 
$\eta_{2H}$ and $\eta_{2V}$ in Supplementary Fig.~\ref{figS1:SFGmodel}, 
the minimum detection efficiency allowed for each of Alice's and Bob's detectors
is estimated. It is found that $S>2$ holds for $\eta_{1H}=\eta_{1V}=\eta_{2H}=\eta_{2V}=0.68$, 
which is much smaller than 0.911 which is the minimum detection efficiency allowed by LO-BSA~\cite{Tsujimoto_2020}. 
These results show the qualitative difference between SFG-BSA and LO-BSA. 

Finally, we estimate the extent to which the SFG efficiency must be improved such that the heralded quantum state exhibits correlations sufficient to violate the Bell-CHSH inequality~($S>2$) and to yield a positive key rate in DIQKD.
In this simulation, we use the experimental parameters in Supplementary Table~\ref{tableS1:Parameters} 
and the dark count probability of D1 as $R_d\tau_W=6.7\times10^{-11}$. 
In our current system, the dark count probability is still much higher than the heralding probability~($\mathrm{Tr}[\hat{\rho}\u{SFG}]=7.6\times10^{-12}$), 
which implies that most of the heralding signal is caused by the dark counts in D1. 
In this situation, the normalized quantum state heralded by the click signal from D1 is given by 
\begin{equation}
\hat{\rho}\u{herald}=\frac{\hat{\rho}\u{SFG}+R_d\tau_W(\tr_{aH,aV,bH,bV}[\ketbra{\psi_{in}}{\psi_{in}}])}{\tr[\hat{\rho}\u{SFG}+R_d\tau_W\ketbra{\psi_{in}}{\psi_{in}}]}, 
\label{eqS4:rhoh}
\end{equation}
where $\tr_{aH,aV,bH,bV}[\ketbra{\psi_{in}}{\psi_{in}}]$ is the uncorrelated quantum state heralded by a dark count. 
We substituted the experimental parameters in Supplementary Table~\ref{tableS1:Parameters}, $R_d=0.15$ and $\tau_W=448\times10^{-12}$ 
for Supplementary Eq.~(\ref{eqS4:rhoh}), and optimized the measurement angles. 
Assuming $\eta_{1H}=\eta_{1V}=\eta_{2H}=\eta_{2V}=1$, we obtain $S=1.88<2$, which indicates that the quantum state heralded by our SFG-BSA would not violate the Bell-CHSH inequality. Nevertheless, it is revealed that $S>2$ would be achieved if the SFG efficiency is tripled. 
To apply for DIQKD, further improvement of the SFG efficiency and optimization of the average photon numbers are necessary, because not only $\chi(S)$ in Supplementary Eq.~(\ref{eq:keyrate}) but also $h(Q)$ must be much smaller than 1. In the absence of loss in the SFG-BSA, i.e. $\eta_t\eta_d=1$, the key rate becomes positive if the SFG efficiency increases by a factor of 50. When the loss in the SFG-BSA is taken into account, an improvement of approximately 140 times in SFG efficiency is required. These levels of improvement are considered realistic in light of recent progress in nonlinear resonators reporting efficiency improvements of two to three orders of magnitude compared to PPLN/W~\cite{Lu:20,Akin2024-cg}.


\end{document}